\documentclass[12pt]{iopart}
\usepackage{epsfig}   
\usepackage{graphics}

\def\half{{1\over 2}}

\def\etal{{\it et~al.\ }}
\def\eg{{\it e.g.,~}}
\def\ie{{\it i.e.~}}

\def\ltsima{$\; \buildrel < \over \sim \;$}
\def\simlt{\lower.5ex\hbox{\ltsima}}
\def\gtsima{$\; \buildrel > \over \sim \;$}
\def\simgt{\lower.5ex\hbox{\gtsima}}

\def \lleq {\lower0.9ex\hbox{ $\buildrel < \over \sim$} ~}
\def \ggeq {\lower0.9ex\hbox{ $\buildrel > \over \sim$} ~}

\def \lam    {\Lambda}
\def\rhoc{\rho_{0{\rm c}}}

\def \om   {\Omega_m}
\def \omt  {\Omega_{0 {\rm m}}}
\def \ol   {\Omega_{\Lambda}}
\def\oc {\Omega_{\kappa}}

\def \beq  {\begin{equation}}
\def \eeq  {\end{equation}}
\def \ber  {\begin{eqnarray}}
\def \eer  {\end{eqnarray}}

\def\apj{{Astroph.\@ J.\ }}
\def\mn{{Mon.\@ Not.\@ Roy.\@ Ast.\@ Soc.\ }}

\def\aj{{Astron.\@ J.\ }}
\def\prl{{Phys.\@ Rev.\@ Lett.\ }}
\def\pd{{Phys.\@ Rev.\@ D\ }}

\def \jetpl {JETP Lett.\ }

\begin{document}

\title{Rejoinder to ``No Evidence of Dark Energy Metamorphosis'',
astro-ph/0404468}
\author{
Ujjaini Alam${}^{a}$, Varun Sahni${}^{a}$, Tarun Deep Saini${}^{b}$ 
and A. A. Starobinsky${}^{c}$}

\address{${}^{a}$ Inter-University Centre for Astronomy \& Astrophysics,
Pun\'e 411 007, India}
\address{${}^{b}$ Institute of Astronomy, Madingley Road, Cambridge, UK}
\address{${}^{c}$ Landau Institute for Theoretical Physics,
119334 Moscow, Russia}
\date{\today}

\begin{abstract}
In a recent paper (astro-ph/0311364) Alam \etal argued that the SNe
data of Tonry \etal 2003 and Barris \etal 2003 appear to favour DE
which evolves in time, provided no other priors are invoked.  (The
effect of invoking priors such as the age of the Universe, the values
of $H_0$ and $\omt$ and CMB/LSS observations could modify this
conclusion, as demonstrated in astro-ph/0403687 and other recent
papers.)  The approach adopted by Alam \etal to reconstruct the
properties of DE was severely (and, as we shall show below --
unfairly) criticized by J\"onsson \etal in astro-ph/0404468.  In this
paper we re-examine the parametrisation used in astro-ph/0311364 and
show that, contrary to the claims of J\"onsson \etal, the results
obtained from this reconstruction are robust and therefore
representative of the true nature of dark energy.
\end{abstract}

\section{Introduction }

The Universe appears to be accelerating and the nature of dark energy
(DE), which drives this acceleration, is a subject of much current
debate among cosmologists. The cosmological constant is the simplest
possibility, but evolving dark energy models have also been suggested
and one of the goals of current cosmological studies is to
differentiate between these different types of models.  Type Ia
supernovae treated as standardized candles provided the first
indications that the expansion of the universe is accelerating
\cite{hzt,perl} and one expects that the nature of dark energy will be
further revealed by the study of these objects in conjunction with
other data sets (CMB, LSS etc.).

Recently there have been several data releases from the two supernova
teams \cite{tonry,barris,knop,riess}, making the total number of
supernovae almost double that known previously. Using these datasets,
we reconstructed the dark energy density and equation of state in two
papers \cite{alam:nov03,alam:mar04} (hereafter Papers I and II) in an
attempt to understand the nature of dark energy. In these works, we
found that the current supernova data appears to favour an evolving
dark energy model with $w \lleq -1$ at present and that at $2\sigma$,
the evolving model is at least as probable as the cosmological
constant. This result soon found support in other works (see for
instance \cite{wang,leandros,huterer,gong1,gong2,gong3,daly,coras}
etc.).  Recently, in \cite{goobar}, doubts have been expressed as to
the reliability of the fitting procedure used in Papers I and II.
Hence in this work we re-examine the reconstruction process used in
these two papers to check whether the results obtained in Papers I \&
II were merely a consequence of the analysis or whether they do
actually reflect a property of the universe.

\section{Reconstructing Dark Energy from Supernova Data}

The SNe observations measure the luminosity distance $d_L(z)$.  From
this, information about the cosmological parameters may be obtained
through the Hubble parameter, which in a spatially flat universe is
related to the luminosity distance quite simply by
\cite{st98,turner,nak99}
\beq
H(z) = \left[ \frac{d}{dz} \left( \frac{d_L(z)}{1+z} \right) \right]^{-1}.
\eeq

The dark energy density is defined as :
\beq\label{eq:dens}
\rho_{\rm DE}=\rhoc \left[\left(\frac{H}{H_0}\right)^2-\omt (1+z)^3\right]\,\,,
\eeq
where $\rhoc=3 H^2_0/(8 \pi G)$ is the present day critical density of
an FRW universe, and $\omt$ is the present day matter density with
respect to the critical density.

Information extracted from SNe observations regarding $d_L(z)$
therefore translates directly into knowledge of $H(z)$, the dark
energy density, and, through \cite{saini00}
\ber\label{eq:state}
q(x) &=& - \frac{\ddot{a}}{a H^2} \equiv \frac{H^{\prime}}{H} x -1 ~, \\
w(x) &=& {2 q(x) - 1 \over 3 \left( 1 - \om(x) \right)}
\equiv \frac{(2 x /3) \ H^{\prime}/H - 1}{1 \ - \ (H_0/H)^2
\omt \ x^3} ~;~x=1+z\,\,,
\eer
into knowledge about the deceleration parameter of the universe and
the equation of state of dark energy.

One route to the reconstruction of DE lies in the construction of an
ansatz for one of the three quantities: $H(z), d_L(z)$ or $w(z)$. The
ansatz must of course be sufficiently versatile to accommodate a large
class of DE models. In Papers I and II we have used a polynomial fit
to dark energy density of the form :
\beq
h(x) =  \frac{H(x)}{H_0} =
\left\lbrack \omt x^3 + A_0 + A_1x + A_2 x^2\right\rbrack^\half\,\, , \ x= 1+z\,\,,
\label{eq:taylor}
\eeq
where $A_0+A_1+A_2 = 1-\omt$. Note that this ansatz should not be
considered as a truncated Taylor series for $h^2(z)$.  Rather, it is
an interpolating fit for $h^2(z)$ having the right behaviour for small
and large values of $z$. The number of terms in this fit is sufficient
given the amount and accuracy of the present supernovae data.  With
more and better data in the future, more terms with intermediate (\eg,
half-integer) powers of $x$ may be added to it.

This is equivalent to the following ansatz for DE density (with
respect to the critical density) : 
\beq
\tilde \rho_{\rm DE}(x) = \rho_{\rm DE}/\rhoc = A_0 + A_1x + A_2 x^2\,\,,
\eeq
which is exact for the cosmological constant $w = -1$ ($A_1 = A_2 =
0$) and for DE models with $w = -2/3$ ($A_0 = A_2 = 0$) and $w = -1/3$
($A_0 = A_1 = 0$).

The corresponding expression for the equation of state of DE is :
\beq
\label{eq:state1}
w(x)=-1+\frac{A_1 x+2 A_2 x^2}{3(A_0+A_1 x+A_2 x^2)}\,\,.
\eeq

The likelihood for the parameters of the ansatz can be determined by
minimising a $\chi^2$-statistic:
\beq
\chi^2(H_0,\omt,p_j)=\sum_i \frac{[\mu_{{\rm fit},i}(z_i;H_0,\omt,p_j)-\mu_{0,i}]^2}{\sigma^2_i}\,\,,
\eeq
where $\mu_{0,i}=m_B-M=5{\rm log}d_L+25$ is the extinction corrected
distance modulus for SNe at redshift $z_i$, $\sigma_i$ is the
uncertainty in the individual distance moduli (including the
uncertainty in galaxy redshifts due to a peculiar velocity of $400$
km/s), and $p_j$ are the parameters of the ansatz used. We assume a
flat universe for our analysis but make no further assumptions on the
nature of dark energy. We also marginalise over the nuisance parameter
$H_0$ by integrating the probability density ${\rm e}^{-\chi^2/2}$
over all values of $H_0$.

\begin{figure*}
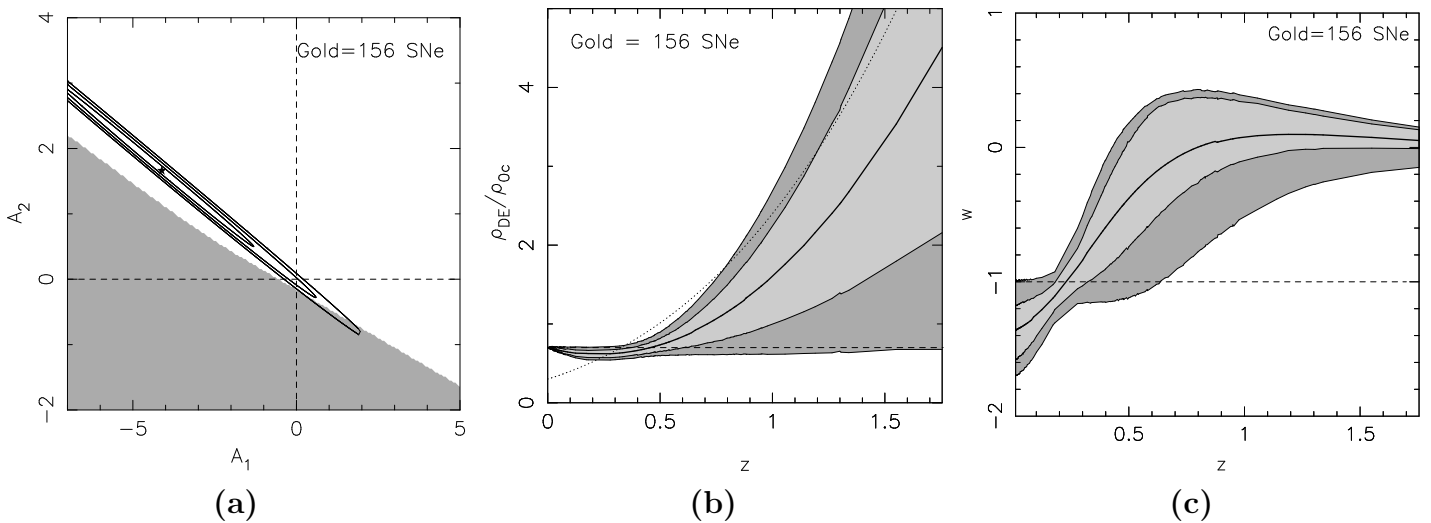

\centering
\begin{center}
\vspace{0.05in}
$\begin{array}{@{\hspace{-0.5in}}c@{\hspace{0.1in}}c@{\hspace{0.1in}}c}
\multicolumn{1}{l}{\mbox{}} &
\multicolumn{1}{l}{\mbox{}} &
\multicolumn{1}{l}{\mbox{}} \\ [-0.5cm]
\epsfxsize=2.4in
\epsffile{h_exp_ell_g.epsi} &
\epsfxsize=2.4in
\epsffile{h_exp_dens_g.epsi} &
\epsfxsize=2.4in
\epsffile{h_exp_w_g.epsi} \\
\mbox{\bf (a)} & \mbox{\bf (b)} & \mbox{\bf (c)}
\end{array}$
\end{center}
\caption{\small
Results from analysis of the 156 SNe ``Gold'' data from \cite{riess},
using ansatz~(\ref{eq:taylor}). $\omt=0.3$ and $h$ is marginalised
over. Panel (a) shows the $1\sigma,2\sigma,3\sigma$ confidence levels
in the $(A_1,A_2)$ parameter space. The star marks the best-fit and
the $\lam$CDM point is at the intersection of the dashed lines. The
shaded grey region has $\rho_{\rm DE} \leq 0$ in the redshift range $0
< z \lleq 2$. Panel (b) shows the variation of dark energy density
$\rho_{\rm DE}(z)/\rhoc$ (where $\rhoc=3 H_0^2/8 \pi G$ is the present
critical energy density) with redshift. Panel (c) shows the evolution
of dark energy equation of state with redshift. In both the panels (b)
and (c), the thick solid line represents the best-fit, the light grey
contours represent the $1\sigma$ confidence level, and the dark grey
contours represent the $2\sigma$ confidence levels. The dashed line in
panels (b) and (c) represents $\lam$CDM, and the dotted line in panel
(b) represents the matter density.}
\label{fig:h_exp_g}
\end{figure*}

\begin{figure*}
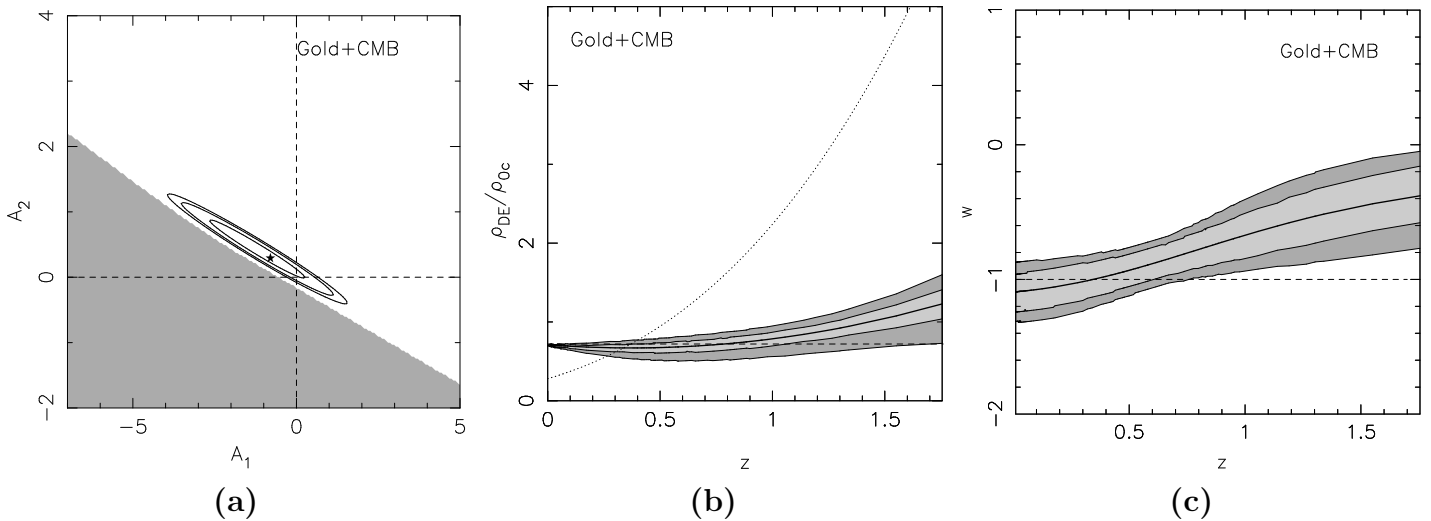

\centering
\begin{center}
\vspace{0.05in}
$\begin{array}{@{\hspace{-0.5in}}c@{\hspace{0.1in}}c@{\hspace{0.1in}}c}
\multicolumn{1}{l}{\mbox{}} &
\multicolumn{1}{l}{\mbox{}} &
\multicolumn{1}{l}{\mbox{}} \\ [-0.5cm]
\epsfxsize=2.4in
\epsffile{h_exp_ell_g_cmb_constr.epsi} &
\epsfxsize=2.4in
\epsffile{h_exp_dens_g_cmb_constr.epsi} &
\epsfxsize=2.4in
\epsffile{h_exp_w_g_cmb_constr.epsi} \\
\mbox{\bf (a)} & \mbox{\bf (b)} & \mbox{\bf (c)}
\end{array}$
\end{center}
\caption{\small
Results for analysis of SNe(Gold)+CMB data with $\lam$CDM-based WMAP
priors of $\omt=0.27 \pm 0.04$ and $h=0.71 \pm 0.06$, using
ansatz~(\ref{eq:taylor}). Panel (a) shows the
$1\sigma,2\sigma,3\sigma$ confidence levels in the $(A_1,A_2)$
parameter space. The star marks the best-fit and the $\lam$CDM point is
at the intersection of the dashed lines.  The shaded grey region has
$\rho_{\rm DE} \leq 0$ in the redshift range $0 < z \lleq 2$.  Panel
(b) shows the variation of dark energy density $\rho_{\rm
  DE}(z)/\rhoc$ (where $\rhoc=3 H_0^2/8 \pi G$ is the present critical
energy density) with redshift. Panel (c) shows the evolution of dark
energy equation of state with redshift. In both the panels (b) and
(c), the thick solid line represents the best-fit, the light grey
contours represent the $1\sigma$ confidence level, and the dark grey
contours represent the $2\sigma$ confidence levels. The dashed line in
panels (b) and (c) represents $\lam$CDM, and the dotted line in panel
(b) represents the matter density.}
\label{fig:h_exp_g_cmb_constr}
\end{figure*}

Figure~\ref{fig:h_exp_g} shows the results obtained by using the
ansatz~(\ref{eq:taylor}) for the ``Gold'' dataset of 156 SNe published
in \cite{riess}. We find that the cosmological constant is just within
the $2\sigma$ confidence level in the $A_1-A_2$ plane. The variation
of dark energy density and equation of state of dark energy is also
shown, and both show indications of evolution. We note that these
results are stable to the addition of extra terms such as $x^4$ or
$x^{-1}$ terms to the ansatz. The region where the dark energy density
becomes less than zero for the redshift range $0 < z \lleq 2$ (the
redshift range in which current supernova data is available) is shown
as the shaded grey region in figure~\ref{fig:h_exp_g}a. We see that a
portion of the $3\sigma (99\%)$ confidence level lies in the region
where dark energy density goes to zero and therefore $w$ blows up.

Figure~\ref{fig:h_exp_g_cmb_constr} shows the results of
reconstruction when results from the WMAP experiment are incorporated
and the priors $\omt=0.27 \pm 0.04$ and $h=0.71 \pm 0.06$ are assumed.
In this case the time-evolution of DE is considerably weaker and more
in agreement with a cosmological constant.  (Note however that
evolving DE is preferred over a cosmological constant by the combined
SNe+WMAP data if $\omt \simeq 0.4, h \simeq 0.6$ as shown in Paper
II, \cite{alam:mar04}.)

\section{Examining the reconstruction exercise for possible bias }

\begin{figure}
\begin{center}
\vspace{-0.05in}
\epsfxsize=3.4in
\epsffile{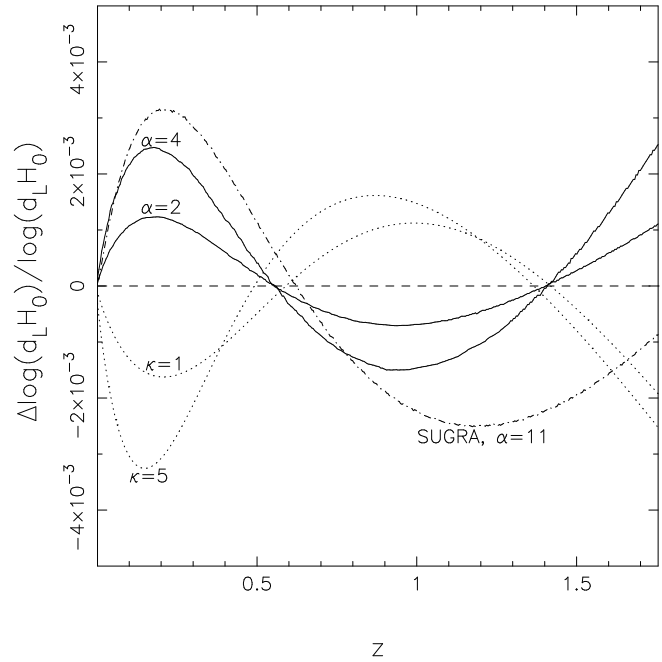}
\end{center}
\caption{\small
The fractional deviation $\Delta{\rm log}(d_L H_0)/{\rm log}(d_L H_0)$
between actual value and that calculated using the
ansatz~(\ref{eq:taylor}) over redshift for different models of dark
energy with $\omt=0.3$. The solid lines represent quintessence tracker
models for potential $V=V_0/\phi^{\alpha}$, with $\alpha=2$ and $4$.
The dotted lines show the deviation for Chaplygin Gas models with
$\kappa=1$ and $5$ (where $\kappa$ is the ratio between CDM and
Chaplygin gas densities at the commencement of the matter dominated
epoch).  The dot-dashed line represents the SUGRA potential,
$V=\left(M^{4+\alpha}/\phi^{\alpha}\right) {\rm
  exp}[\frac{1}{2}\left(\phi/M_{Pl}\right)^2]$, with $M=1.6\times
10^{-8} M_{Pl}, \alpha=11$. The dashed horizontal line represents zero
deviation from model values, which is true for $\lam$CDM and $w=-1/3,
w=-2/3$ quiessence models.  }
\label{fig:err}
\end{figure}

We now look at the reconstruction exercise in some detail to
understand whether the results obtained in Papers I and II are biased
due to inherent weaknesses in the ansatz used or whether these results
can be trusted to reflect the true nature of dark energy.

\subsection{Reconstruction of different dark energy models }

In \cite{goobar}, the ansatz~(\ref{eq:taylor}) was criticised for
being biased and giving misleading information about the nature of
dark energy. We show here that this is not the case and the ansatz is
sufficiently robust to be a reliable probe of the nature of dark
energy.  In the figure~\ref{fig:err}, we show the {\em maximum
  deviation} between the actual value of the luminosity distance and
that calculated using the ansatz~(\ref{eq:taylor}) for simulations of
different physically well motivated models of dark energy such as
quintessence, Chaplygin gas, and the SUGRA potential.  From
(\ref{eq:taylor}) it is clear that for $w=-1$, $w=-2/3$, and $w=-1/3$,
this ansatz returns exact values. However, even for models for which
it is not exact, the ansatz (\ref{eq:taylor}) still recovers the
luminosity distance to within $0.5\%$ accuracy in the redshift range
relevant for SNe observations.  These results reassure us that for a
large class of models, the ansatz can be trusted to recover
cosmological quantities to a high degree of accuracy.  (Other tests of
the ansatz (\ref{eq:taylor}) have been reported in
\cite{sahni03,alam:mar03,alam:nov03}.)

\subsection{Diverging equation of state }

In \cite{goobar} the ansatz (\ref{eq:taylor}) has been criticized on
the grounds that it allows the DE density to become negative during
the course of cosmological evolution.  We would like to emphasise that
this is not necessarily a bad thing since there is no a-priori reason
for the DE density to remain positive throughout its evolution and
models do exist in which $\rho_{\rm DE} \leq 0$ for a substantial
duration of time -- see for instance
\cite{felder,frieman,kallosh1,kallosh2,ass03} etc.  Indeed the purpose
of the construction of an ansatz such as (\ref{eq:taylor}) is to endow
it with sufficient flexibility so that it is able to reproduce the
behaviour of a sufficiently large class of DE models.\footnote{It is
  clear that no finite parameter ansatz will will ever be able to
  reproduce all possible behaviour of DE.  For instance the ansatz
  (\ref{eq:taylor}) is unlikely to give good results if applied to DE
  which evolves extremely rapidly with time such as the rapidly
  oscillating models of \cite{sahni_wang}.}  In addition, a negative
value for the dark energy density could also be an indication that the
matter density has been over estimated, so one should not arbitrarily
exclude the region of parameter space where $\rho_{\rm DE} < 0$ as
suggested by \cite{goobar}.  Hence we have not a priori put a
constraint of positivity on the dark energy density in our analysis.
However, one point to note is that the region where dark energy
density vanishes is very close to the cosmological constant
($A_1=A_2=0$) in parameter space, so there is a chance that if the
best-fit to the data is near this point then the best-fit $w$ or
errors in it may blow up.  This need not always happen, for instance
in the figure~\ref{fig:h_exp_g_cmb_constr}, the best-fit is close to
the cosmological constant, but the errors on $w$ are well-behaved. The
same is seen in figure 6 of Paper I.  However, in case the best-fit
equation of state or the errors in it do blow up, we naturally have to
be careful about whether this truly indicates some exotic form of dark
energy, or whether it is more consistent with the cosmological
constant.  It is also important to point out that the equation of
state parameter need not always be the best characteristic of dark
energy \cite{alam:mar03}, and in such cases it is useful to
characterize DE using other cosmological parameters (which do not
exhibit this divergence), such as the deceleration parameter and the
Statefinder pair \cite{sahni03,alam:mar03}.  In this context it is
worth noting that the pressure of dark energy : $P/\rhoc=-\rho_{\rm
  DE}+A_1 x/3+2 A_2 x^2/3$ is well-behaved for this ansatz even when
$\rho_{\rm DE} < 0$, and this quantity can therefore be used to draw
conclusions on the nature of dark energy.

\subsection{Non-monotonic errors on the equation of state }

\begin{figure}
\centering
\begin{center}
\vspace{-0.05in}
\epsfxsize=2.4in
\epsffile{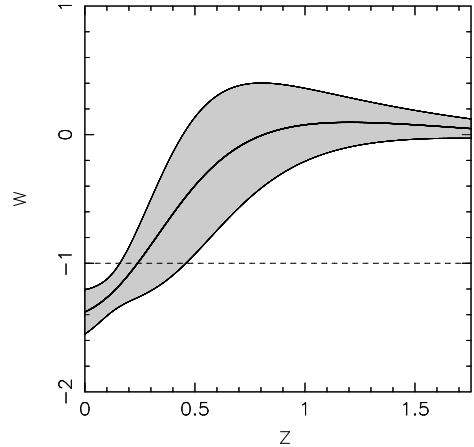}
\end{center}
\caption{\small
The variation of the equation of state of dark energy $w(z)$ over
redshift for the ansatz~(\ref{eq:taylor}). The thick solid line shows
the best-fit and the light grey contour represents the $1\sigma$
confidence level around the best-fit.  The dashed horizontal line
denotes $\lam$CDM. $\omt = 0.3$ is assumed. Errors are calculated by the
Fisher matrix approach, using Eq~\ref{eq:err_w}.}
\label{fig:err_w}
\end{figure}

The errors on the equation of state in figure~\ref{fig:h_exp_g}(c)
appear to be non-monotonic. This seems somewhat counter-intuitive. Low
redshift behaviour of the equation of state affects the luminosity
distance at all higher redshifts, while high redshift behaviour
affects fewer points. This leads to an expectation that high-$z$
behaviour of the equation of state should be poorly constrained as
opposed to the low-$z$ behaviour. However, we saw in Papers I \& II
that the errors in $w(z)$ actually decrease with redshift. We
investigate this matter using the Fisher matrix error bars.

In an analysis which uses an ansatz with $n$ parameters $p_i$, the
Fisher information matrix is defined to be 
\beq
F_{ij} \equiv \left \langle \frac{\partial^2 {\bf L}}{\partial p_i \partial p_j} \right \rangle\,\,,
\eeq
where ${\bf L}=-{\rm log}\cal{L}$, $\cal{L}$ being the likelihood .
For an unbiased estimator, the errors on the parameters will follow
the Cram\'{e}r-Rao inequality : $\Delta p_i \geq 1/\sqrt{F_{ii}}$.

Since the likelihood function is approximately Gaussian near the
maximum likelihood (ML) point, the covariance matrix for a maximum
likelihood estimator is given by
\beq
(C^{-1})_{ij} \equiv \frac{\partial^2 {\bf L}}{\partial p_i \partial p_j}\,\,.
\eeq
The Fisher information matrix is therefore simply the expectation
value of the inverse of the covariance matrix at the ML-point.
  
Given the covariance matrix, the error on any cosmological quantity
$Q(p_i)$ is given by :
\beq\label{eq:propagate}
\sigma_Q^2 = \sum_{i=1}^n \left( \frac{\partial Q}{\partial p_i} \right)^2 C_{ii}+2 \sum_{i=1}^n \sum_{j=i+1}^n  \left( \frac{\partial Q}{\partial p_i} \right)  \left( \frac{\partial Q}{\partial p_j} \right) C_{ij}\,\,.
\eeq
Thus the nature of the errors on a quantity will depend essentially on
the manner in which it is related to the parameters of the system.

The errors on the equation of state for the ansatz (\ref{eq:taylor})
can now be calculated using equations~(\ref{eq:taylor})
and~(\ref{eq:propagate}), and has the somewhat complicated expression :
\beq\label{eq:err_w}
\sigma_w^2(x) = \frac{x^2 [f_1^2 C_{11} + 2 f_1 f_2 C_{12}+ f_2^2 C_{22}]}{9 [1-\omt+A_1 (x-1)+A_2 (x^2-1)]^4}\,\,,
\eeq
where 
\begin{eqnarray*}
f_1 &=& 1-\omt-A_2 (x-1)^2 ~, \\
f_2 &=& 2 x (1-\omt)+A_1 (x-1)^2 \,\,,
\end{eqnarray*}
and $A_1,A_2$ are the best-fit values of the parameters.  

From this expression it is difficult to predict whether the errors
should increase with redshift or not. Indeed that would depend on the
value and sign of the quantities $C_{ii}$.  We calculate the errors
around the best-fit for the 156 SNe ``Gold'' dataset of \cite{riess}
using this procedure that the $1\sigma$ confidence levels thus
calculated on the equation of state (shown in figure~\ref{fig:err_w})
are almost identical to those shown in figure~\ref{fig:h_exp_g}(c).
\footnote{One should note that the Fisher matrix approach works best
  for the cases where the likelihood is symmetric around the best-fit.
  For asymmetric likelihood curves, this method should not be used.}

Thus the ansatz~(\ref{eq:taylor}) need not show monotonically
increasing errors in $w$ with redshift and in this way may differ from
other fitting functions, for instance approximations to the equation
of state of dark energy as proposed in \cite{linder} or
\cite{bassett}, which may show just the opposite tendency. This
indicates that the nature of error bars are affected by the quantity
being approximated. In the limit of infinite terms in the expansion of
various quantities all the methods should produce identical result.
The practical need for truncating these expansions make these
approximations slightly different from each other. More specifically,
we require setting of priors
\ber
f(z) &=& \sum_{i=0}^{\infty} a_n z^n\\
a_n &=& 0 ; \ (n> N_p)
\eer
where $f(z)$ could be $H(z)$, $w(z)$ or any other physical quantity
and $N_p$ is the chosen number of parameters. The non-linear priors in
the above equation make different finite expansions inequivalent.
Since we do not know for certain if the underlying model for the
accelerating expansion involves an energy component with negative
pressure in a FRW setting we are forced to choose one of the
alternatives for approximations. We hope that with increasingly high
quality data the effect of such truncations will eventually disappear.

\subsection{Stability of the dark energy equation of state }

The dark energy equation of state for the ansatz~(\ref{eq:taylor}) can
have many different forms depending on the values of $A_1,A_2$. This
ansatz gives us an exact representation of the cosmological constant
($A_1=A_2=0$) and it is an unbiased estimator for the $\lam$CDM model.
However, it is also true that the statistical noise inherently present
in the data will never allow us to measure $A_1,A_2$ to be exactly
equal to zero even in this case.  Thus if the underlying model of the
universe is the cosmological constant, then by fitting an ansatz which
has more parameters than required, we are fitting a bit of noise as
well. This fact is true for {\em any} ansatz and not just
(\ref{eq:taylor}).  The deviation of the best-fit from the
cosmological constant in such a case is given by :
\beq
\delta w_{\Lambda} \simeq \frac{x}{3(1-\omt)}\delta A_1 + \frac{2 x^2}{3(1-\omt)}\delta A_2 \,\,,
\eeq 
\ie the $w_{\Lambda}$ curve would increasingly depart from the
cosmological constant with redshift. The same tendency would be seen
if we consider a linear ansatz in the equation of state : $w=w_0+w_1
z$. This would also diverge from the cosmological constant ($w_0 = -1,
w_1 = 0$) at high redshifts for even a small non-zero $\delta w_1$,
since $\delta w_{\Lambda} \simeq \delta w_0 + \delta w_1 z$.  Such
divergence is the inevitable price we pay for not knowing the ``true''
nature of dark energy and for using an ansatz to explore it. In
practice one is free to choose either of two possibilities: (i) assume
that the underlying model has certain characteristics (such as
constant equation of state) and thus face the danger of losing
information-- the pitfalls of using this approach were highlighted in
\cite{maor}, or (ii) we may choose to make no such assumptions and
consider models with a somewhat larger number of parameters. We have
adopted (ii) in Papers I \& II and believe that, given the redshift
range and statistical noise of the data currently available, it is not
unreasonable to assume that the errors for the
ansatz~(\ref{eq:taylor}) will not be large enough to cause the
best-fit to diverge significantly from the underlying model
irrespective of whether the model is that of the cosmological constant
or some other form of dark energy. This is demonstrated in figure
\ref{fig:sim_h_exp_dens}(a) for simulations of the cosmological
constant model.

\subsection{Correlation of the parameters $A_1,A_2$ }

\begin{figure}
\begin{center}
\vspace{-0.05in}
\epsfxsize=2.4in
\epsffile{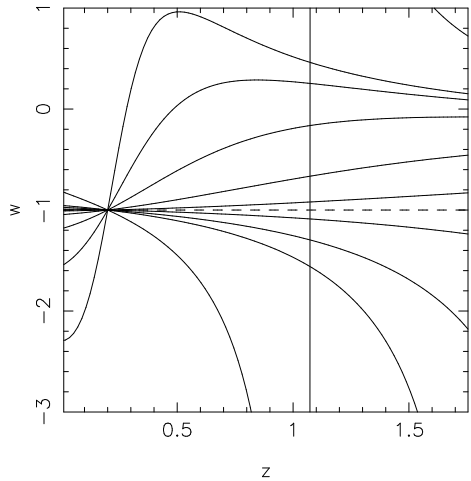}
\end{center}
\caption{\small
The variation of the equation of state with redshift for different
values of the parameters $A_1,A_2$ satisfying $A_1=-2.4A_2$. The thick dashed
line denotes $\lam$CDM.}
\label{fig:h_exp_correl}
\end{figure}

In our analysis of the supernova data we find that for a fixed value
of $\omt$, the parameters $A_1,A_2$ which are favoured by the data are
strongly correlated. For $\omt=0.3$, from the confidence levels of
figure~\ref{fig:h_exp_g}(a) we may conclude that approximately, $A_1
\simeq -2.4 A_2$ for points favoured by the data.  Contrary to what is
claimed in \cite{goobar} this correlation is not something that has
been put into the analysis by hand, rather it is a consequence of the
observational data. Other fits, such as the Linder fit to the equation
of state \cite{linder}, have also shown a strong correlation between
its parameters. We need to examine what effect, if any, this
correlation has on the nature of dark energy. From expression
(\ref{eq:state1}) for the equation of state for dark energy, we see
that $w=-1$ at a redshift of $z=-A_1/2A_2-1$.  Therefore, for
$\omt=0.3$, $w=-1$ at $z \simeq 0.2$. This implies that since most
curves would cross $w=-1$ at approximately this redshift, there will
be a ``sweet-spot'' at $z \simeq 0.2$ for the equation of state. This
can be clearly seen in the figure~\ref{fig:h_exp_g}(c).  The presence
of sweet-spots at different redshifts for different ansatz is
something which has been studied earlier, for example in
\cite{albrecht}, and the community agrees that the presence or absence
of a sweet-spot is simply a consequence of the ansatz used and does
not signify any extra information about the underlying model. Does the
fact that $w$ crosses $-1$ at this redshift mean that the ansatz
forces the equation of state to evolve in a particular direction? In
figure~\ref{fig:h_exp_correl}, we plot the behaviour of $w$ for
different values of the parameters $A_1,A_2$ which follow the
constraint $A_1=-2.4 A_2$. We see that the cosmological constant is
allowed by this constraint, as are many other models. One can have
extremely mild evolution of $w$, or more rapid evolution of $w$, one
can have $w$ which blows up, $w$ which goes from a more negative
($<-1$) value to a less negative ($>-1$) value or vice versa. In short
this constraint does not force the equation of state to evolve at a
particular rate in a particular direction, but leaves a great deal of
freedom. Also, it is important to note that this is not a constraint
that has been imposed by hand in the analysis, it is simply an outcome
of the analysis. There are many points in the $A_1,A_2$ parameter
space for which the equation of state is always $\geq-1$ and shows
little evolution. The ansatz does not preclude these points, it is the
data which chooses a particular evolving dark energy model over other
possibilities. We therefore conclude that the correlation seen in the
parameters $A_1,A_2$ does not signify any bias in the results and
(contrary to the claims of \cite{goobar}) the ansatz does not
``force'' evolution of the dark energy equation of state.

\subsection{Confirmation from other fitting functions }

In Paper I, we showed that the results obtained by the
ansatz~(\ref{eq:taylor}) can also be reproduced using other fitting
functions, such as the fits to the equation of state used in
\cite{linder} and \cite{bassett}. For both these fits, the equation of
state shows an evolution which is remarkably similar to that found
using (\ref{eq:taylor}).  It is interesting that our results have
found support in the independent analysis carried out by other groups
which also show that evolving DE provides a fit that is as good (or
better) than that provided by a cosmological constant to the current
supernova data \cite{wang,leandros,huterer,gong1,gong2}. Therefore
there seems to be a consensus among the different groups that, based
on the SNe data alone, an evolving dark energy model is a plausible
alternative to the cosmological constant.  (As emphasised earlier in
this paper and in Papers I and II, the properties of dark energy
depend crucially upon the manner in which the SNe data is sampled and
also upon which other sets of observations are used in conjunction
with type Ia supernovae.  For instance, the evolution of DE is least
if one uses SNe+WMAP results together with the priors $\omt = 0.3,
h = 0.7$; see also \cite{bagla}.)

\section{Simulation of a $\lam$CDM model }

\begin{figure}
\begin{center}
\vspace{-0.05in}
\epsfxsize=6.0in
\epsffile{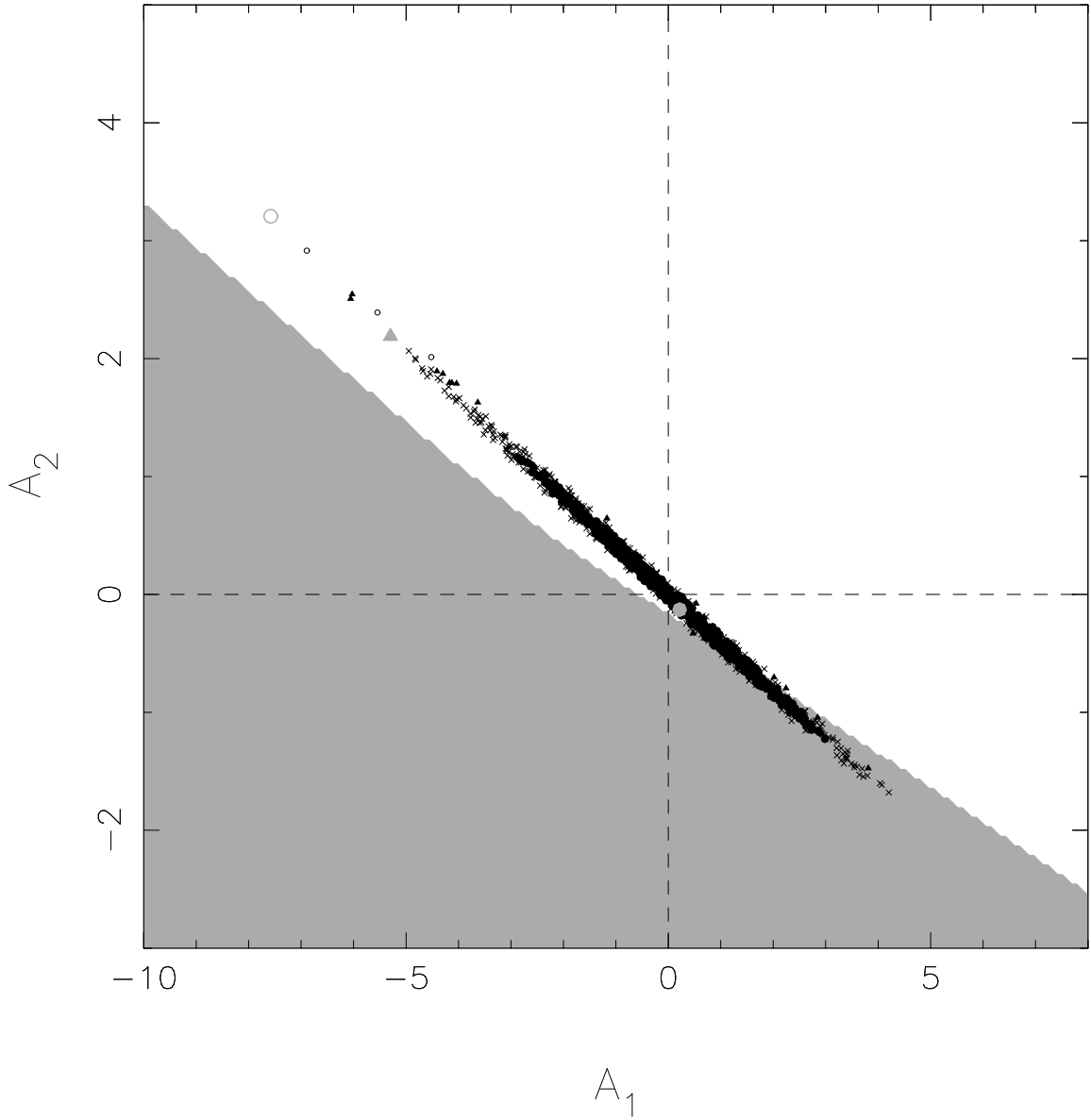}
\end{center}
\caption{\small
A scatter plot of the best-fit values $A_1,A_2$ of the polynomial fit
to DE density, Eq~(\ref{eq:taylor}), for 1000 simulated data sets in a
flat universe with a cosmological constant. $\omt=0.3$ is assumed. The
filled circles represent simulated datasets consistent with $\lam$CDM at
the $68\%$ confidence level, the crosses represents those consistent
at the $95\%$ confidence level, the filled triangles are consistent at
the $99\%$ confidence level, while the open circles are inconsistent
at the $99\%$ confidence level. Three of these datasets used in
further analysis are shown by (a) a grey circle, (b) a grey triangle,
and (c) a grey open circle. The grey shaded region has $\rho_{\rm DE}
\leq 0$ in the redshift range $0 < z \lleq 2$.}
\label{fig:sim_h_exp_ell}
\end{figure}

\begin{figure*}
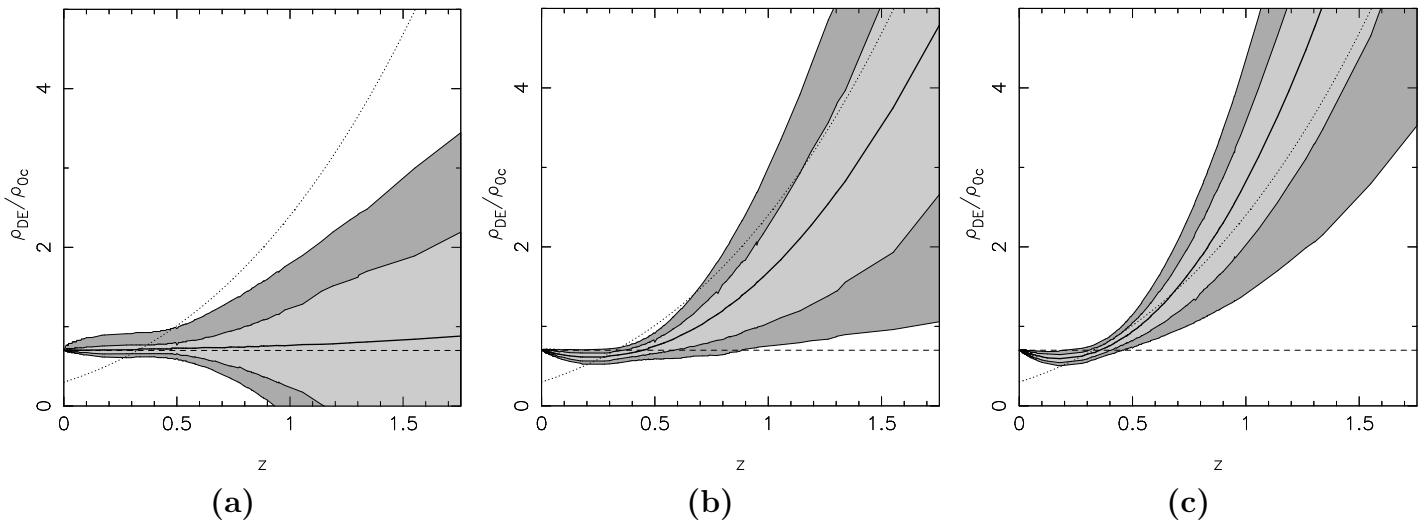

\centering
\begin{center}
\vspace{0.05in}
$\begin{array}{@{\hspace{-0.5in}}c@{\hspace{0.1in}}c@{\hspace{0.1in}}c}
\multicolumn{1}{l}{\mbox{}} &
\multicolumn{1}{l}{\mbox{}} &
\multicolumn{1}{l}{\mbox{}} \\ [-0.5cm]
\epsfxsize=2.4in
\epsffile{h_exp_dens_sngl_1.epsi} &
\epsfxsize=2.4in
\epsffile{h_exp_dens_sngl_4.epsi} &
\epsfxsize=2.4in
\epsffile{h_exp_dens_sngl_5.epsi} \\
\mbox{\bf (a)} & \mbox{\bf (b)} & \mbox{\bf (c)}
\end{array}$
\end{center}
\caption{\small
The variation of dark energy density $\rho_{\rm DE}(z)/\rhoc$ (where
$\rhoc=3 H_0^2/8 \pi G$ is the present critical energy density) with
redshift, using three sets of simulated data for a fiducial flat
$\lam$CDM model. The polynomial fit to DE density, Eq~(\ref{eq:taylor})
with $\omt=0.3$ in a flat universe has been used.  In each panel, the
thick solid line shows the best-fit, the light grey contour represents
the $1\sigma$ confidence level, and the dark grey contour represents
the $2\sigma$ confidence level around the best-fit.  The dotted line
denotes matter density $\omt (1+z)^3$, and the dashed horizontal line
denotes $\lam$CDM. }
\label{fig:sim_h_exp_dens}
\end{figure*}

\begin{figure*}
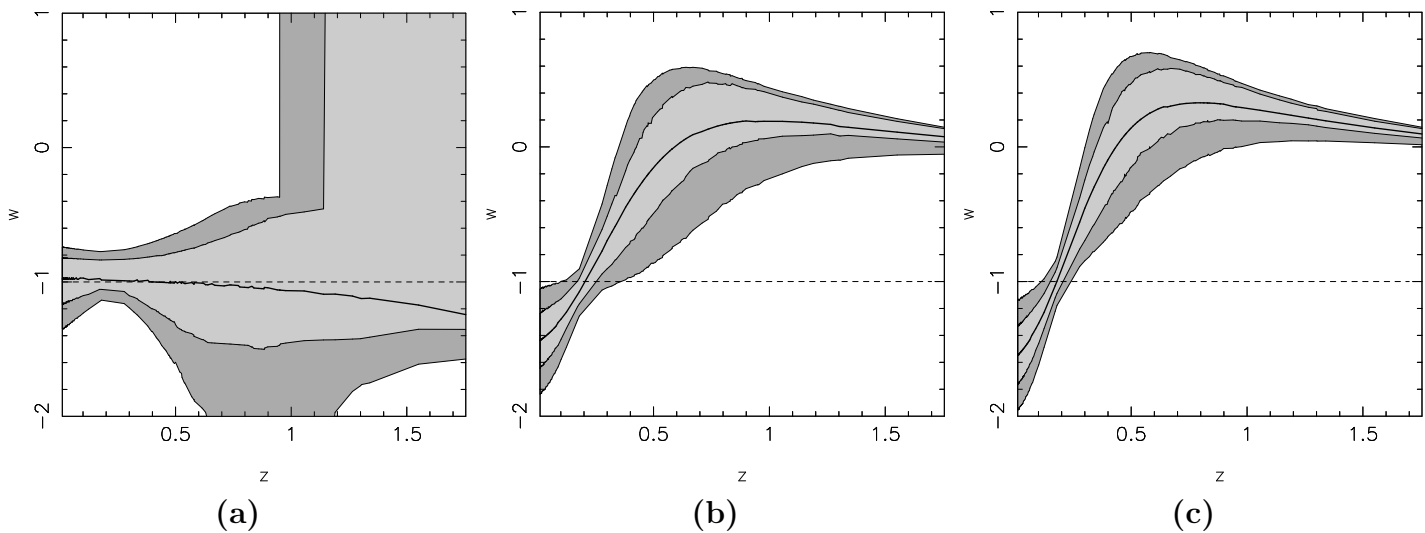

\centering
\begin{center}
\vspace{0.05in}
$\begin{array}{@{\hspace{-0.5in}}c@{\hspace{0.1in}}c@{\hspace{0.1in}}c}
\multicolumn{1}{l}{\mbox{}} &
\multicolumn{1}{l}{\mbox{}} &
\multicolumn{1}{l}{\mbox{}} \\ [-0.5cm]
\epsfxsize=2.4in
\epsffile{h_exp_w_sngl_1.epsi} &
\epsfxsize=2.4in
\epsffile{h_exp_w_sngl_4.epsi} &
\epsfxsize=2.4in
\epsffile{h_exp_w_sngl_5.epsi} \\
\mbox{\bf (a)} & \mbox{\bf (b)} & \mbox{\bf (c)}
\end{array}$
\end{center}
\caption{\small
The variation of equation of state of dark energy $w(z)$ with redshift
for $\omt=0.3$ in a flat universe using three sets of simulated data
for the polynomial fit to DE density, Eq~(\ref{eq:taylor}). The
background model is flat $\lam$CDM. In each panel, the thick solid line
shows the best-fit, the light grey contour represents the $1\sigma$
confidence level, and the dark grey contour represents the $2\sigma$
confidence level around the best-fit.  The dashed horizontal line
denotes $\lam$CDM.}
\label{fig:sim_h_exp_w}
\end{figure*}

To test the reliability of our ansatz, we attempt to reconstruct a
fiducial cosmology from simulated data. We assume that the background
model is a flat $\lam$CDM universe with $\omt=0.3$, and generate 1000
datasets, each consisting of 156 SNe with the same redshift
distribution and magnitude errors as those in the ``Gold'' dataset of
\cite{riess}. We first analyse these datasets using the polynomial fit
to dark energy density, Eq~(\ref{eq:taylor}). The best-fit values of
$A_1,A_2$ for these 1000 sets are shown in
figure~\ref{fig:sim_h_exp_ell}. The mean of this set of $1000$ points
is very close to the cosmological constant: $\bar{A_1}=-0.01,
\bar{A_2}=0.005$ ($A_1 = A_2 = 0$ for $\lam$CDM).  The best-fit points
consistent with the cosmological constant at $68\%$ confidence level
are indicated by filled circles, points consistent at $95\%$
confidence level are shown by crosses, those consistent at $99\%$
confidence level are shown by filled triangles, and those inconsistent
at $99\%$ confidence level are shown by the open circles. By
determining confidence levels around the best-fit for each individual
dataset we find that about $75\%$ of the best-fit points are
consistent with the cosmological constant at $1\sigma$.  We also note
that some of the best-fit points are in the region where $\rho_{DE}
\leq 0$. We now choose three of these 1000 datasets and calculate
$\rho_{DE}$ and $w$ for the corresponding cosmology. The three
best-fit points chosen are shown-- the first (hereafter called dataset
(a)) is consistent within $68\%$ confidence level, the next (dataset
(b)) is consistent at $99\%$ confidence level and the last (dataset
(c)) is inconsistent at $99\%$ confidence level with the cosmological
constant. The results are shown in figures~\ref{fig:sim_h_exp_dens}
and~\ref{fig:sim_h_exp_w}. We find that for the dataset (a),
$\rho_{DE}$ and $w$ are consistent with the cosmological constant,
even though the errors in $w$ blow up. For the datasets (b) and (c),
the nature of dark energy density and dark energy equation of state
appear to be very different from that of the cosmological constant.
Such an example was used in figure 4 of \cite{goobar} to make the
claim that the ansatz~(\ref{eq:taylor}) leads us to conclude that $w$
is evolving even though the fiducial model is that of the cosmological
constant.

\begin{figure*}
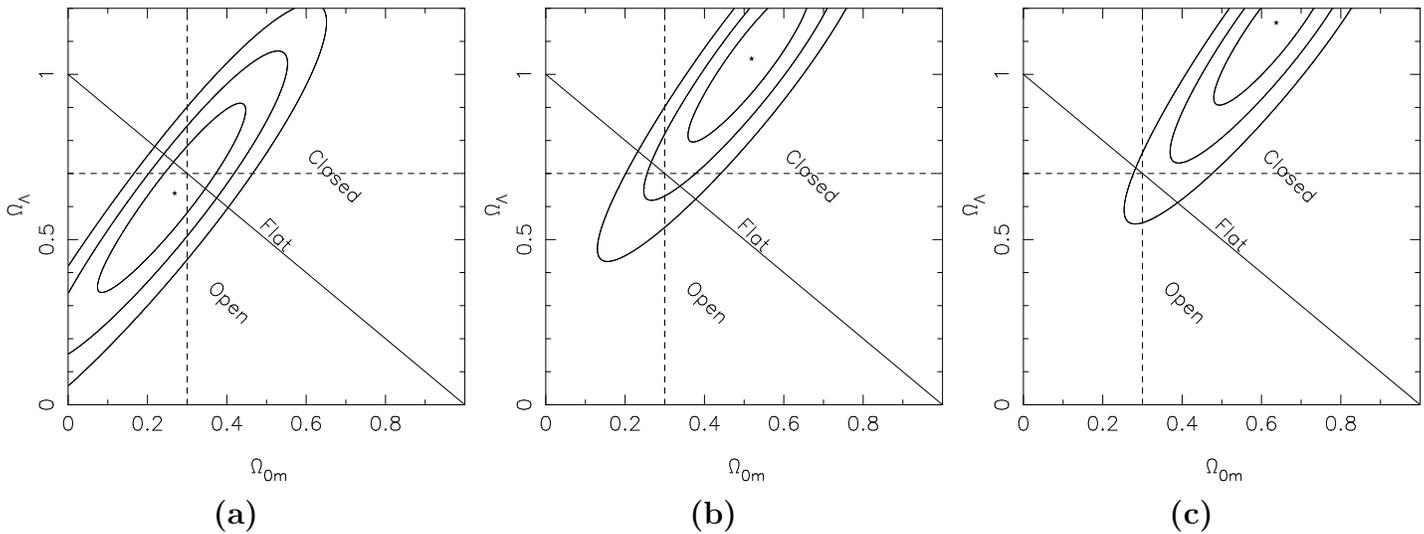

\centering
\begin{center}
\vspace{0.05in}
$\begin{array}{@{\hspace{-0.5in}}c@{\hspace{0.1in}}c@{\hspace{0.1in}}c}
\multicolumn{1}{l}{\mbox{}} &
\multicolumn{1}{l}{\mbox{}} &
\multicolumn{1}{l}{\mbox{}} \\ [-0.5cm]
\epsfxsize=2.4in
\epsffile{om_ol_ell_sngl_1.epsi} &
\epsfxsize=2.4in
\epsffile{om_ol_ell_sngl_4.epsi} &
\epsfxsize=2.4in
\epsffile{om_ol_ell_sngl_5.epsi} \\
\mbox{\bf (a)} & \mbox{\bf (b)} & \mbox{\bf (c)}
\end{array}$
\end{center}
\caption{\small
The $1\sigma,2\sigma,3\sigma$ confidence levels in $\omt,\ol$ for the
three sets of simulated data (with a flat cosmological constant
background model, $\omt + \ol = 1$) using a cosmological constant
model, Eq~(\ref{eq:om_ol}) for the analysis.  In each panel, a star
marks the best-fit.  The intersection of the dashed lines denotes
$\lam$CDM. The solid line denotes the flat universe with $\omt + \ol =
1$. All universes above this line are closed and those below it are
open. }
\label{fig:sim_om_ol}
\end{figure*}

Let us now see if this is really correct and whether our results were
indeed a construct of the ansatz used.
\begin{itemize}

\item 
Let us consider the same fiducial model ($\lam$CDM) but an alternative
ansatz to reconstruct dark energy.  As our first example we consider a
cosmological constant model but make no assumption about the flatness
of the universe. The luminosity distance in this case is given by :
\ber
&&\hspace{-1.5cm}d_L(z)=\frac{c(1+z)}{H_0 \sqrt{|\oc|}} {\cal S}\left\lbrace \sqrt{|\oc|} \int_0^z\frac{dz}{\sqrt{(1+z)^2(1+\omt z)-z(2+z)\ol}}\right\rbrace \label{eq:om_ol}\\
&&(\oc=1-\omt-\ol; {\cal S}={\rm sin,1,sinh} \ {\rm for} \ \oc <,=,> 0) \nonumber\,\,.
\eer
The parameters of the system are $\omt, \ol$. The results are shown in
the figure~\ref{fig:sim_om_ol}. We see that in case of dataset (a),
the errors on $\omt, \ol$ are completely consistent with the
cosmological constant in a flat universe, but for both the datasets
(b) and (c) a closed model is the preferred model and the flat model
(which is the true background model) is excluded at $1\sigma$ and
$2\sigma$ respectively !

We may also consider different fitting functions for $w$ in a flat
universe.

\item 
Consider the popular linear fitting function for the equation of state :
\ber
w(z)&=&w_0+w_1 z\label{eq:w_riess}\\
H^2(z)&=&H_0^2[\omt (1+z)^3+(1-\omt)(1+z)^{3 (1+w_0-w_1)}{\rm e}^{3 w_1 z}]\nonumber\,\,.
\eer
This is the ansatz used in \cite{riess} to parametrise dark energy.
The resultant confidence levels in $w_0, w_1$ are shown in the
figure~\ref{fig:sim_w_riess}. Once again, we see that only in case of
the dataset (a) does the parametrisation come close to reconstructing
the actual model. For the dataset (b), the cosmological constant is
ruled out at $1\sigma$ and for the dataset (c) it is ruled out at
$2\sigma$.

\begin{figure*}
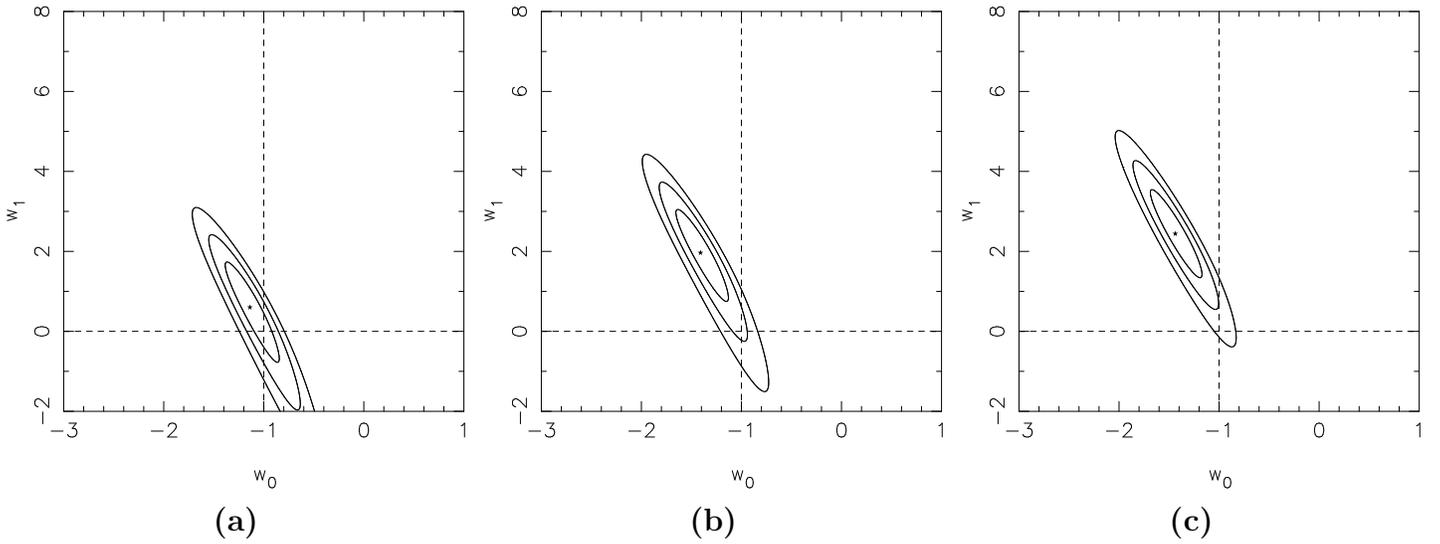

\centering
\begin{center}
\vspace{0.05in}
$\begin{array}{@{\hspace{-0.5in}}c@{\hspace{0.1in}}c@{\hspace{0.1in}}c}
\multicolumn{1}{l}{\mbox{}} &
\multicolumn{1}{l}{\mbox{}} &
\multicolumn{1}{l}{\mbox{}} \\ [-0.5cm]
\epsfxsize=2.4in
\epsffile{w_riess_ell_sngl_1.epsi} &
\epsfxsize=2.4in
\epsffile{w_riess_ell_sngl_4.epsi} &
\epsfxsize=2.4in
\epsffile{w_riess_ell_sngl_5.epsi} \\
\mbox{\bf (a)} & \mbox{\bf (b)} & \mbox{\bf (c)}
\end{array}$
\end{center}
\caption{\small
The $1\sigma,2\sigma,3\sigma$ confidence levels in $w_0,w_1$ using
three sets of simulated data for the linear fit to equation of state
of DE, Eq~(\ref{eq:w_riess}). $\omt=0.3$ and a flat universe are
assumed. Background model is flat $\lam$CDM. In each panel, a star marks
the best-fit.  The intersection of the dashed lines denotes $\lam$CDM. }
\label{fig:sim_w_riess}
\end{figure*}

\begin{figure*}
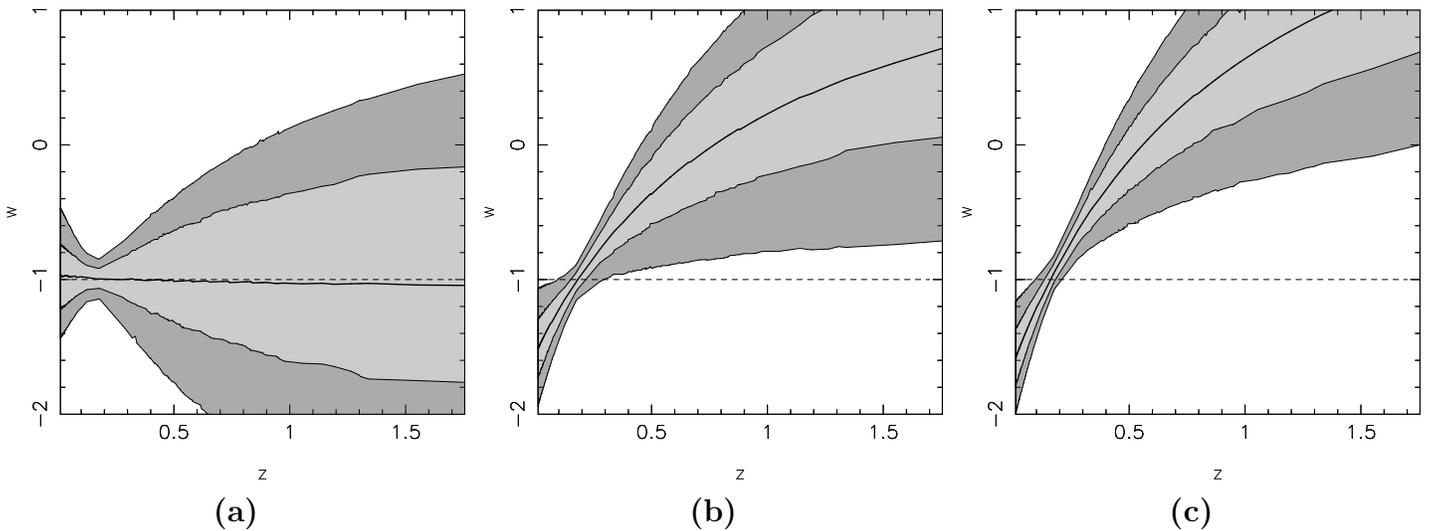

\centering
\begin{center}
\vspace{0.05in}
$\begin{array}{@{\hspace{-0.5in}}c@{\hspace{0.1in}}c@{\hspace{0.1in}}c}
\multicolumn{1}{l}{\mbox{}} &
\multicolumn{1}{l}{\mbox{}} &
\multicolumn{1}{l}{\mbox{}} \\ [-0.5cm]
\epsfxsize=2.4in
\epsffile{w_lin_w_sngl_1.epsi} &
\epsfxsize=2.4in
\epsffile{w_lin_w_sngl_4.epsi} &
\epsfxsize=2.4in
\epsffile{w_lin_w_sngl_5.epsi} \\
\mbox{\bf (a)} & \mbox{\bf (b)} & \mbox{\bf (c)}
\end{array}$
\end{center}
\caption{\small
The variation of equation of state of dark energy $w(z)$ with redshift
for $\omt=0.3$ and a flat universe using three sets of simulated data
for Linder's fit to the equation of state of DE, Eq~(\ref{eq:w_lin}).
Background model is flat $\lam$CDM. In each panel, the thick solid line
shows the best-fit, the light grey contour represents the $1\sigma$
confidence level, and the dark grey contour represents the $2\sigma$
confidence level around the best-fit.  The dashed horizontal line
denotes $\lam$CDM.  }
\label{fig:sim_w_lin}
\end{figure*}

\item 
Finally consider the parametrisation suggested in \cite{linder} :
\ber
w(z)&=&w_0+\frac{w_1 z}{1+z}\label{eq:w_lin}\\
H^2(z)&=&H_0^2[\omt (1+z)^3+(1-\omt)(1+z)^{3 (1+w_0+w_1)}{\rm e}^{-3 w_1 z/(1+z)}]\nonumber\,\,.
\eer
The evolution of the equation of state with redshift in this case for
the three datasets is shown in the figure~\ref{fig:sim_w_lin}. We see
that in the first panel (a), the evolution is very moderate and is
entirely consistent with a cosmological constant. In the second and
third panels however, we see significant evolution of the equation of
state, which is consistent with what is seen in the panels (b) and (c)
of figure~\ref{fig:sim_h_exp_w}, and inconsistent with a cosmological
constant model.

\end{itemize}

The above examples clearly indicate that the different fits give
results which are mutually consistent and that, contrary to the claims
made in \cite{goobar}, the ansatz~(\ref{eq:taylor}) does not ``force''
evolution of the equation of state of dark energy when the data is
commensurate with the cosmological constant. If the datasets for which
(\ref{eq:taylor}) shows evolution of DE are analysed using other fits,
similar (misleading) conclusions about the behaviour of DE are
reached. So these results are dependent not so much on the ansatz as
on the dataset itself.  Therefore it is in principle possible that
even with a flat $\lam$CDM fiducial model having errors commensurate
with current observations one can reconstruct dark energy which
evolves, or a universe that is closed ! However, one must remember
that out of the $1000$ realisations of such a model, about $75\%$ are
broadly consistent with the cosmological constant model for the
ansatz~(\ref{eq:taylor}) and if these datasets are analysed then the
correct model will be recovered within $1\sigma$, as in case of the
dataset (a). The datasets (b) and (c) which give rise to incorrect
results are much less likely to occur. As the quality of data
improves, the likelihood of obtaining datasets like (b) and (c) will
further decrease. Given the current observational standards, it is
reasonable to assume that the actual data which we have is consistent
with the underlying model and results obtained from it are reliable.
As more data becomes available, we will be able to discern the nature
of the universe with greater accuracy.

\section{Conclusion }

We have shown that the results of Papers I and II are robust and that
the conclusions drawn there are not due to any bias in the ansatz
used.  Contrary to the claims in \cite{goobar} the
ansatz~(\ref{eq:taylor}) {\em does not} inherently ``force'' or
``favour'' evolving dark energy models over the cosmological constant.
This ansatz can reproduce the behaviour of a large class of DE models
fairly accurately and as such it can be used with confidence to
predict the nature of the universe using currently available data. The
conclusion that the current supernova data favours the evolving dark
energy models over the cosmological constant at $1\sigma$ still holds
and this result has found support in the work of other independent
groups \cite{wang,leandros,huterer,gong1,gong2,gong3,daly,coras} when
no other data sets are used.  Better quality data expected in the
future from different cosmology experiments (SNe, CMB, LSS etc.) will
allow us to draw firmer conclusions about the nature of dark energy.

\end{document}